%% file: paper.tex
\documentclass[10pt,conference]{IEEEtran}
\IEEEoverridecommandlockouts
\usepackage[table]{xcolor}
\usepackage{cite}
\usepackage{amsmath,amssymb,amsfonts}
\usepackage{algorithmic}
\usepackage{graphicx}
\usepackage{textcomp}
\usepackage{subcaption}
\usepackage{xcolor}
\usepackage[utf8]{inputenc}
\usepackage{balance}
\input{macros}

\def\BibTeX{{\rm B\kern-.05em{\sc i\kern-.025em b}\kern-.08em
    T\kern-.1667em\lower.7ex\hbox{E}\kern-.125emX}}
\usepackage{booktabs} 

\usepackage{tabularx}

\begin{document}

\title{Automatic Static Bug Detection for Machine Learning Libraries: Are We There Yet?}

\author{
    \IEEEauthorblockN{Nima Shiri harzevili\IEEEauthorrefmark{1},
     Jiho Shin\IEEEauthorrefmark{1}, 
    Junjie Wang\IEEEauthorrefmark{2},
    Song Wang\IEEEauthorrefmark{1},
   Nachiappan Nagappan\IEEEauthorrefmark{3}}
    \IEEEauthorblockA{\IEEEauthorrefmark{1}York University; \IEEEauthorrefmark{2}Institute of Software, Chinese Academy of Sciences;
    \IEEEauthorrefmark{3}META
     \\\{nshiri,jihoshin,wangsong\}@yorku.ca; junjie@iscas.ac.cn; nachiappan.nagappan@gmail.com}
}
\maketitle

\begin{abstract}
	\input{sec/abstract}
\end{abstract}

\begin{IEEEkeywords}
Software bugs, static detection, machine learning libraries
\end{IEEEkeywords}

\input{sec/introduction}
\input{sec/motivation}
\input{sec/approach}

\input{sec/result}

\input{sec/lessons}

\input{sec/validity}
\input{sec/related}
\input{sec/conclusion}

\balance
\bibliographystyle{IEEEtran}
\bibliography{paper}

\end{document}

%% file: macros.tex
\usepackage{soul}
\usepackage{hyperref}
\usepackage{multirow}
\usepackage{fancybox}
\usepackage{enumitem}
\usepackage{graphicx}
\usepackage{color}
\usepackage{float}
\usepackage{xcolor}
\usepackage{array}
\usepackage{amsmath}
\usepackage{xspace}
\usepackage{url}
\usepackage{tikz}
\usepackage{caption}
\usepackage{pgfplots}
\usepackage{listings}
\usepackage{color}
\usepackage{graphicx}
\definecolor{dkgreen}{rgb}{0,0.6,0}
\definecolor{gray}{rgb}{0.5,0.5,0.5}
\definecolor{mauve}{rgb}{0.58,0,0.82}
\pgfplotsset{compat=1.16}
\usepackage{pgf-pie}
\usetikzlibrary{pgfplots.statistics,calc}
\usepackage{algorithm}
\usepackage{algorithmic}
\usepackage{subcaption}
\usepackage{listings}

\usepackage{tcolorbox}

\makeatletter
\newcommand{\mybox}[1]{%
	\setbox0=\hbox{#1}%
	\setlength{\@tempdima}{\dimexpr\wd0+13pt}%
	\begin{tcolorbox}[boxrule=0.5pt, colback=gray!10, arc=4pt,
		left=6pt,right=6pt,top=6pt,bottom=6pt,boxsep=0pt]
		#1
	\end{tcolorbox}
}

\definecolor{codegreen}{rgb}{0,0.6,0}
\definecolor{codegray}{rgb}{0.5,0.5,0.5}
\definecolor{codepurple}{rgb}{0.58,0,0.82}
\definecolor{backcolour}{rgb}{0.95,0.95,0.92}

\lstdefinestyle{mystyle}{
  language=Python,
  aboveskip=3mm,
  showstringspaces=false,
  columns=flexible,
  numbers=none,
  backgroundcolor=\color{backcolour},
  commentstyle=\color{codegreen},
 keywordstyle=\color{magenta},
    numberstyle=\tiny\color{codegray},
    stringstyle=\color{codepurple},
    basicstyle=\small\ttfamily,
    breakatwhitespace=false,         
    breaklines=false,                 
    captionpos=b,                    
    keepspaces=false,                 
    numbersep=5pt,                  
    showspaces=false,                
    showstringspaces=false,
    showtabs=false,                  
    tabsize=2,
    escapeinside=``
}
\newcommand{\ctext}[3][RGB]{%
  \begingroup
  \definecolor{hlcolor}{#1}{#2}\sethlcolor{hlcolor}%
  \hl{#3}%
  \endgroup
}

%% file: sec/abstract.tex
Automatic detection of software bugs is a critical task in software security. Many static tools that can help detect bugs have been proposed. While these static bug detectors are mainly evaluated on general software projects call into question their practical effectiveness and usefulness for machine learning libraries. In this paper, we address this question by analyzing five popular and widely used static bug detectors, i.e., Flawfinder, RATS, Cppcheck, Facebook Infer, and Clang static analyzer on a curated dataset of software bugs gathered from four popular machine learning libraries including Mlpack, MXNet, PyTorch, and TensorFlow with a total of 410 known bugs. Our research provides a categorization of these tools’ capabilities to better understand the strengths and weaknesses of the tools for detecting software bugs in machine learning libraries. Overall, our study shows that static bug detectors find a negligible amount of all bugs accounting for 6/410 bugs (0.01\%), Flawfinder and RATS are the most effective static checker for finding software bugs in machine learning libraries. Based on our observations, we further identify and discuss opportunities to make the tools more effective and practical. 

%% file: sec/introduction.tex
\section{Introduction}
\label{sec:introduction}

Programming inevitably involves dealing with bugs in software, which is an aspect that can be frustrating for many developers since detecting and fixing bugs is time-consuming~\cite{foreman2019vulnerability, madry2017towards}. To help developers find software bugs, many static bug detectors have been developed and are now frequently employed by many industries and open-source  projects~\cite{cheng2021deepwukong, cao2022mvd, li2021vuldeelocator, bessey2010}. Error Prone from Google~\cite{errorprone}, Infer from Facebook \cite{infer}, and SpotBugs~\cite{SpotBugs}, the successor to the widely used FindBugs tool \cite{ayewah2008using}, are examples of popular static bug detection applications. These tools are usually developed as an analytical framework based on static analysis and are capable of scaling to large applications. 

Previous research has examined static bug detectors on traditional projects from different aspects~\cite{johnson2013don, thung2012extent, thung2015extent, habib2018many, tomassi2021real, lipp2022empirical}. {There are two major limitations in existing studies. First, the datasets used for the empirical evaluation of static bug detectors are not real-world examples, they are not able to replicate new and sophisticated bug patterns. Second, they do not categorize the type of bugs based on Common Weakness Enumeration(CWE) information. 
Recently, Lipp et al.~\cite{lipp2022empirical} addressed the limitations by proposing an empirical evaluation of static bug detectors on real-world datasets collected from CVE records gathered from 27 projects containing 1.15 million lines of code.} Their results showed that state-of-the-art tools can detect in-between 20\% and 53\% of the bugs in a benchmark set of real-world programs. 
However, it has not been determined if the findings of these studies on conventional projects are applicable to machine learning (ML) libraries. Finding real-world security bugs in ML libraries is critical for a couple of reasons. First, ML libraries have been widely used in many fields in the past decades, such as image classification \cite{algan2021image, mahdisoltani2018fine}, big data analysis \cite{patgiri2018taxonomy}, pattern recognition \cite{lv2022semi}, autonomous driving \cite{simhambhatla2019self, ramos2017detecting, kulkarni2018traffic}, and natural language processing \cite{minaee2017automatic, athreya2021template, roy2021deep}. Failure to detect bugs in these ML libraries might have devastating implications, such as traffic accidents~\cite{hong2020artificial}. 
Second, gaining an understanding of the benefits and drawbacks of the static bug detectors that are currently in use will direct future work toward the design of new methodologies or tools for detecting ML-specific bugs.

In this paper, we set out to investigate the effectiveness of five popular static bug detectors on ML libraries. Specifically, we conduct an empirical analysis utilizing 410 real-world bugs collected from four widely-used popular ML libraries to address the issue of how many bugs from these ML libraries can be detected by static bug detectors {and why they miss detecting real-world security bugs of ML libraries}. We select five popular and open-source static bug detectors as the research subjects, i.e., Flawfinder~\cite{flawfinder, chen2023memory, lipp2022empirical, pereira2020use, mitropoulos2019employing, gu2022hierarchical, zou2022mvulpreter, cheng2021deepwukong, zhou2019devign}, RATS~\cite{rats, zou2022mvulpreter, cheng2021deepwukong}, Cppcheck~\cite{cppcheck, chen2023memory, cheng2021deepwukong, zhou2019devign}, Facebook Infer~\cite{infer, pujar2022varangian, mehrpour2023can, mehrpour2023can}, and Clang static analyzer~\cite{lattner2008llvm, cheng2021deepwukong, umann2022detecting, szecsi2022improved, aslanyan2022static}. The fundamental strategy entails employing each bug detector on a version of a program that is afflicted with a particular bug and determining whether or not the bug is discovered by the bug detector.
The primary obstacle of the proposed empirical study is determining whether or not the collection of warnings that a bug detector has reported actually captures the issued bug. To tackle the problem, we filtered the reported warnings based on their line number that involve in the bug fix following existing studies~\cite{pradel2011detecting}.  
Then we conduct a manual inspection of the detected bug candidates to avoid possible coincidental matches. 
In addition, we have also investigated bug detectors given a set of program analysis criteria such as input representation, pattern matching techniques, and source code element sensitivity. Specifically, we manually examined the documentation as well as the source code of these static bug detectors to understand how they function. 
We also study the characteristics of vulnerabilities in ML libraries and understand their patterns. As a result, we provide future directions to improve existing static checkers to deal with specific vulnerabilities in ML libraries.

This paper has the following contributions: 
\begin{itemize}
    \item We present the first empirical study on the investigation of current static bug detection on four widely-used machine learning libraries.
    \item We identify gaps between existing static bug detection techniques and software security practices on machine learning libraries.
    \item We explore the capabilities and limitations of each static bug detector, as well as suggest future ideas for enhancing the effectiveness of these tools in real-world settings to detect software vulnerabilities specific to ML libraries. 
    \item We release the dataset of our experiments to help other researchers replicate and extend our study\footnote{\tiny{https://anonymous.4open.science/r/ISSRE2023SATS-E255/README.md}}.
\end{itemize}

\begin{figure*}[t!]
\centering
\begin{subfigure}{\columnwidth}
    \begin{lstlisting} [escapechar=`,
                        numbers=left,
                        firstnumber=43,
                        language = Python,
                        escapeinside={(*}{*)}]
  gtl::InlinedVector<npy_intp, 4> dims(ndims);
  if (TF_TensorType(tensor) == TF_RESOURCE) {
    dims[0] = TF_TensorByteSize(tensor);
    CHECK_EQ(ndims, 0)}
        << "Fetching of non-scalar 
        resource tensors is
        not supported.";
    dims.push_back(TF_TensorByteSize(tensor));
    *nelems = dims[0];
  } else {
    *nelems = 1;
    \end{lstlisting} 
    \caption{An example of buffer overflow in TensorFlow library.}
    \label{fig:buffer1}
\end{subfigure}
\begin{subfigure}{\columnwidth}
   \begin{lstlisting} [
                        escapechar=`,
                       language = C,
                       escapeinside={(*}{*)}]
chip = get_chip_info(sdev->pdata);
for (i = 0; i < HDA_EXT_ROM_STATUS_SIZE; i++) {
value = snd_sof_dsp_read(sdev, HDA_DSP_BAR, 
  chip->rom_status_reg + i * 0x4);
		len += snprintf(msg + len, sizeof(msg) - len, 
  " 0x%x", value);
		len += scnprintf(msg + len, sizeof(msg) - len, 
  " 0x%x", value);
	}
dev_printk(level, sdev->dev, 
"extended rom status: %s", msg);
    \end{lstlisting} 
    \caption{An example of buffer overflow in Linux kernel.}
    \label{fig:buffer2}
\end{subfigure}
  \caption{This figure shows two examples of buffer overflow from the TensorFlow library and Linux kernel. }
    \label{fig:cwebuffer}
\end{figure*}

%% file: sec/motivation.tex
\section{Motivation}

ML libraries and traditional software differ in various aspects such as complexity, data dependency, and performance characteristics. Overall, ML libraries tend to be more complex than traditional software as machine learning often involves complex mathematical models and algorithms~\cite{harzevili2022characterizing, thung2012empirical, islam2019comprehensive, jia2021symptoms}. In addition, as ML models are heavily dependent on the quality and quantity of the data used in their training process, most ML libraries provide tools for data pre-processing, cleaning, and normalization that are not available in traditional software. ML libraries are designed to take advantage of modern hardware, including GPUs and specialized processors, to optimize performance. Traditional software, on the other hand, is generally optimized for general-purpose CPUs.




These differences introduce different bugs compared to traditional software. Figure \ref{fig:buffer1} and Figure \ref{fig:buffer2} show two examples of buffer overflow in TensorFlow library\footnote{\tiny{https://github.com/tensorflow/tensorflow/commit/13ef0af4867477cdda7e0b294e61560c2952df42}} and Linux kernel\footnote{\tiny{https://github.com/torvalds/linux/commit/5549af7f42916c0d7e78a0e423ac667e27eaac3e}} respectively. The root cause of buffer overflow in Figure~\ref{fig:buffer1} is lack of proper input validation when getting a non-scalar resource tensor while the root cause of buffer overflow in Figure~\ref{fig:buffer2} is using \textit{snprintf()} instead of using \textit{scnprintf()} as \textit{snprintf()} is not a secure built-in C API. We can see that there are two instances of the same bug (i.e., buffer overflow) that have distinct root causes. To detect a heap buffer overflow example in the TensorFlow library as shown in Figure \ref{fig:buffer1}, existing static bug detectors must follow two crucial steps. Firstly, these tools need to recognize the TensorFlow-specific macro checkers (\textit{CHECK\_EQ}) by incorporating the macro call signature into their internal database of risky APIs. Secondly, they must conduct data flow analysis to trace the flow of data from client APIs to the backend implementation. In this example, the vulnerable parameter is \textit{ndims} which has been coming from the client API usage. Unfortunately, we discovered that the current tools lack the capability to perform these essential steps, rendering them inadequate for detecting machine learning-specific bugs. Please note that there are more complex bug patterns that we discuss in RQ3. 

The different symptoms and root causes of the same type of security vulnerabilities from ML libraries and traditional software motivate us to look deeper into understanding the performance of current static bug detection on machine learning libraries.


%% file: sec/approach.tex
\section{Study design}
\label{sec:methodology}

In this section, we first show how we curated a dataset of real-world bugs of ML libraries (Section~\ref{subsec:data}), then we introduce the static bug detection tools used in this paper (Section~\ref{subsec:detectors}), finally, we describe the procedure of applying bug detectors to our curated real-world data (Section~\ref{subsec:vulIdentification}).

\subsection{Collection of Bugs from ML Libraries}
\label{subsec:data}



Our study is on the basis of four widely used ML libraries selected based on a set of inclusion criteria: 1) libraries should be open source and available to the public, 2) they should be under active development, 3) libraries should have the implementation of classical ML as well as state-of-the-art DL models, 4) libraries should support different tasks in the common ML workflow. The outcome of applying these criteria is the following ML libraries; Mlpack, MXNet, PyTorch, and TensorFlow. Our filters also exclude some well-known libraries including Caffe, Theano, and Keras due to the fact that the data extraction mechanism utilized in this work did not retrieve enough real-world security vulnerabilities from their repositories due to a lack of sufficient vulnerable data in the GitHub repository of these libraries.

For each of the studied ML libraries, we followed the approach proposed by Zhou et al \cite{zhou2017automated} to extract venerability-related commits. Note that we extract all commits in the default branch of each library since the starting date of the development.
Specifically, we use their regular expression rules, including expressions and keywords related to security issues, to collect security bug-fixing commits. As a result, we collected around 5k 
commits to the four studied projects.  
Note that, the collected commits might contain noises due to the fact that there may be coincidental matches between bug keywords and the keywords inside the commit message and title~\cite{zhou2017automated}. 
To remove noises, we further conducted a manual inspection on each commit. 
In our manual inspection, two authors began evaluating extracted commits simultaneously. The authors analyzed the title, message, merged pull requests, and linked issues\footnote{\footnotesize{Some bug fixing commits fix opened issues. Thus, the authors further analyzed them for noise removal.}} of each bug fixing commit and remove the commits that are not related to a registered security bug in CWE website\footnote{https://cwe.mitre.org/}. As a result, we collected 410 bug-fixing commits across the four projects that were investigated shown in Table~\ref{tab:datastat}.








\input{tables/dataStatistics.tex}

\subsection{Running Static Bug Detection Tools}
\label{subsec:detectors}

In this paper, we study five publicly available static bug detection tools. Our primary emphasis is on static detectors that are freely accessible to the public. Please note that three of the tools are under active development including Facebook Infer~\cite{infer}, Clang static analyzer~\cite{lattner2008llvm}, and Cppcheck\footnote{https://cppcheck.sourceforge.io/}. The two remaining tools (Flawfinder \cite{younan2012runtime} and RATS\footnote{https://github.com/andrew-d/rough-auditing-tool-for-security}) are not actively developing but they have been frequently used in software bug detection~\cite{cheng2021deepwukong, li2018vuldeepecker, duan2019vulsniper}. 

The static bug detectors used in this work have several common characteristics. For example, Flawfinder and RATS treat the input source code as a text sequence while Cppcheck, Infer, and Clang static analyzers convert the source code into an intermediate representation. All tools use a built-in database of patterns that are used to detect security vulnerabilities. For instance, RATS, Flawfinder, and Cppcheck all use a database of C/C++ system functions that are known to have security bugs such as buffer overflow or format string issues. while Infer uses a bi-abduction inference and analysis in order to find vulnerable source code statements. Clang static analyzer work by parsing the source code and checking it against a set of predefined checks, or \textit{linters}.

\input{tables/toolcomparison}

For the capabilities of these five static bug detectors used in this paper, following existing work~\cite{tomassi2021real}, we take into consideration the following three different types of program analysis features, i.e., input representation, matching strategies (\textit{intraprocedural} and \textit{interprocedural}), and sensitivity to program elements including \textit{flow}, \textit{context}, \textit{field}, \textit{object}, \textit{path}, and \textit{field}. We investigated the tools in relation to the aforementioned analytical program analysis features. During this procedure, we manually evaluated the tools' source code and documentation, which demonstrated several types of behaviors to check the tools' capabilities and limits. In the following paragraphs, each tool is explained in detail. Table~\ref{tab:toolcomparison} shows the capabilities of static analysis tools investigated.

\textbf{Flawfinder~\cite{flawfinder}}: is a static analysis tool that scans a program for potential security bugs by using a database of known unsafe C/C++ functions. Flawfinder can detect problems with race conditions and system calls in addition to \verb|printf()| and normal string manipulation operations, based on an internal database that contains C/C++ routines that are known to have vulnerabilities in their design. 

\textbf{RATS~\cite{rats}}: An open-source static analyzer that is able to analyze code bases written in C, C++, Perl, PHP, and Python. Similar to Flawfinder, it uses an internal database of risky C/C++ API signatures and uses a keyword-matching approach to find and mark them as vulnerable in the target source code. 

\textbf{Cppcheck~\cite{cppcheck}}: Cppcheck offers one-of-a-kind code analysis to find defects and focuses on finding undefined behavior as well as risky coding structures. The objective is to generate an extremely low number of false positives. 

\textbf{Infer~\cite{infer}}: is a static analysis tool that is developed by Facebook. It searches for a wide variety of vulnerabilities in programs written in Java, C/C++, and Objective-C. Bi-abduction analysis is one of the methods that Infer employs in order to locate vulnerabilities such as deadlocks, memory leaks, and null pointer dereference. In order to perform analysis, Infer requires a set of compilation commands for each file. Infer is an interprocedural analysis tool which means that it allows keeping track of objects and variables between methods, and also global variables.

\textbf{Clang static analyzer~\cite{lattner2008llvm}}: Clang static analyzer is built on top of the LLVM project, which offers a set of modular and reusable compiler and toolchain technologies. It is an extensible framework for C/C++ code linting that may be used to enforce coding standards, conduct static analysis, and discover probable errors. It operates by scanning the source code and comparing it to a collection of predefined checks, known as linters. The checks are built as separate modules, making it simple to add new checks or modify current ones. 

Note that the static bug detectors used in this paper have capabilities to report non-bug related warnings, e.g.,  styling or refactoring issues, as well as reporting general bugs which are not related to software bugs that have a unique id in CWE website\footnote{https://cwe.mitre.org/}. In order to reduce false positive rates and prevent static checkers to produce related warnings to bugs, we have configured them to merely report warnings that are strongly related to software security vulnerabilities. Please note that RATS does not generate styling issues. In addition, we manually studied Flawfinder's built-in database and discovered that all rules have a unique id associated with CWE records. As a result, we solely used this setup for Infer, Cppcheck, and Clang static analyzer. 

\subsection{Identification of bug Candidates}
\label{subsec:vulIdentification}

One of the main challenges in the identification of bug candidates is to check whether the reported warnings by static detectors are corresponding with the issued bug class or not. To identify bug candidates, we use the following steps. First, we performed automatic filtering based on the commits' diff information as used in previous work~\cite{habib2018many,tomassi2021real}. Afterward, we further analyzed the reported warnings that involve the commit using manual inspection.


\subsubsection{Automatic filtering}
\label{sec:mapping}

We filter reported warnings by using a diff-based mapping technique which uses code diffs between vulnerable and fixed programs in a fixing commit \cite{pradel2011detecting}. First, it computes a set of lines in the vulnerable program that is flagged with at least one warning. Then, it checks whether the flagged line numbers overlap with the changed lines in the fixing commit. If the flagged lines overlap with the code change, the warning is considered a bug candidate. Otherwise, it is not a candidate for manual inspection. For example, Figure~\ref{fig:diffmapping} shows an example of a warning generated by Flawfinder that is considered to be a vulnerable candidate since the line number reported in the warning overlaps with the line number in the code change of the bug fixing commit.

\begin{figure*}[t!]
\centering
\begin{subfigure}{\columnwidth}
    \begin{lstlisting}
    for (int i = 0; i < client->device_count(); ++i) {
    se::StreamExecutorConfig config;
    config.ordinal = i; config.device_options.non_portable_tags
    ["host_thread_stack_size_in_bytes"] =
    (* \ctext[RGB]{255, 99, 71}{-absl::StrCat(2048 * 1024);}*)
    (* \ctext[RGB]{0, 153, 76}{+absl::StrCat(8192 * 1024);}*)
    TF_ASSIGN_OR_RETURN(se::StreamExecutor * executor,
    platform->GetExecutor(config));
    \end{lstlisting} 
    \caption{bug fixing commit.}
    \label{fig:diffmapping1}
\end{subfigure}
\begin{subfigure}{\columnwidth}
   \begin{lstlisting} [
                        escapechar=`,
                       language = C,
                       escapeinside={(*}{*)}]
    Examining /cpu_device.cc
    FINAL RESULTS:
    cpu_device.cc:47: 
    [4] (buffer) StrCat:
    Does not check for buffer overflows when
    concatenating to destination
    [MS-banned] (CWE-120).
    absl::StrCat(2048 * 1024) 
    \end{lstlisting} 
    \caption{reported warning by Flawfinder.}
    \label{fig:diffmapping2}
\end{subfigure}
  \caption{An example of bug fixing commit in TensorFlow library in which the developer has increased the thread stack size from 2048 to 8192 to prevent buffer overflow bug. Figure \ref{fig:diffmapping2} shows the generated warning by Flawfinder in which there is a possible buffer overflow at line 47. }
    \label{fig:diffmapping}
\end{figure*}

\subsubsection{Manual verification}

{In this step, we manually scan all candidates and check the warning messages against the vulnerable and clean versions of the code to eliminate potential accidental matches. 
The first two authors examine the warnings and bug-fixing commits simultaneously. They manually analyze the warning message, the CWE-ID, and the line number where the bug is reported. Based on the above information, they review the code changes to confirm whether the warning is related to the bug being addressed in the commit. 
{Note that, in the cases where the detector tools reported a warning at a line outside those changed lines in the bug-fixing commit, we manually checked the warning. In addition, we have also performed a backward analysis in which we traced all variables and function calls within the code changes up to the marked line that was outside the code changes. If the marked line was associated with the code changes, we considered the corresponding warning as a potential bug.} 
If there is any disagreement, they flag the commit and the corresponding warning for further manual verification in the next round. 
They repeat this process several times until all warnings and commits have been reviewed.}

During our manual inspection, we also audit the following information: 1) True Positive (TP): the bug described in the warning precisely matches with the bug reported for the commit; 2) False Positive (FP): the bug reported in the warning does not match with the bug reported in the commit. False Negative (FN): a potentially vulnerable warning mistakenly predicted to be a false alarm. In this paper, we calculate True Positive Rate (TPR) and False Negative Rate (FNR) as $TPR = \frac{TP}{TP+FN}$ and $FNR = \frac{FN}{FN+TP}$ respectively.

%% file: tables/dataStatistics.tex
\begin{table}[t!]
\centering
\caption{Characteristics of bug fixing commits (BFCs) used in this paper.} 
\renewcommand{\arraystretch}{1}
\begin{tabular}{llcc}
\hline
Library & Language & LOC$(\sim)$ & {\# BFCs} \\
\hline
Mlpack     & C++   &  340K  & 47     \\
MXNet  & C++/Python & 362K & 60    \\
PyTorch    & C++/Python & 3.8M & 58   \\
TensorFlow & C++/Python & 567K & 245   \\ \hline
\multicolumn{2}{l}{ Overall}  & 5M & 410 \\
\hline
\end{tabular}
\label{tab:datastat}
\end{table}

%% file: tables/toolcomparison.tex
\begin{table*}[t!]
\centering
\caption{Characteristics of tools based on different program analysis factors. In this table, the asterisk sign indicates their partial capability in considering program analysis factors. Also, \textit{N.A} shows that we do not have enough information about the program analysis factor.}
\label{tab:toolcomparison}
\begin{tabular}{lccccccccccc}
\toprule
\multirow{2}{*}{Tool}& \multirow{2}{*}{Version} & \multicolumn{3}{c}{Input representation} & \multicolumn{2}{c}{Pattern matching} & \multicolumn{5}{c}{Sensitivity}        \\ \cline{3-12} 
&         &Text & AST        & Other        & Intraprocedural   & Interprocedural  & Context & Field & Object & Data Flow & Control Flow \\
\midrule
Flawfinder    & 2.0.19 & $\checkmark$ &  -  & -  & $\times$  & $\times$ &  $\times$  & $\times$   & $\times$   & $\times$  & $\times$  \\
RATS          & 2.4 & $\checkmark$ &  -  & -  &  $\times$ &   $\times$ & $\times$   & $\times$   & $\times$   & $\times$  & $\times$  \\
Cppcheck      & 2.7 & - & $\checkmark$ & - &  $\checkmark$  &   $\times$ &  $\checkmark$ &  $\times$ & $\times$  &  $\ast$ &  $\ast$ \\
Infer         & 1.1.0 & - & - & $\checkmark$ & $\checkmark$ & $\checkmark$ & $\checkmark$  & $\checkmark$ & N.A & $\checkmark$ & $\checkmark$ \\
Clang static analyzer        & 14.0.0 & - & $\checkmark$ & - & $\checkmark$ & $\checkmark$ & $\checkmark$  & $\checkmark$ & $\checkmark$ & $\checkmark$ & $\checkmark$ \\
\bottomrule
\end{tabular}
\end{table*}

%% file: sec/result.tex
\section{Experimental Results}
\label{sec:reAnalysis}
In this section, we present and discuss our analysis results to address the following three research questions.

\noindent \textbf{RQ1 (Detection Capability)}: How many warnings are reported by the studied static bug detectors?
\vspace{4pt}


\noindent \textbf{RQ2 (Detection Effectiveness)}: How effective are static bug detectors at detecting real-world bugs in ML libraries?
\vspace{4pt}

\noindent \textbf{RQ3 (Root Cause)}: What are the root causes of missing real-world bugs in ML libraries?

For answering these RQs, we execute Flawfinder, RATS, Cppcheck, Infer, and Clang static analyzer on our dataset of 410 real-world bugs across the four studied ML libraries. The output of bug detectors, i.e., generated warnings, are automatically parsed to extract relevant information including bug types, line numbers, and warning messages. All data and scripts for conducting the experiments described in this section are accessible to the public. 

\input{sec/results/rq1}

\input{sec/results/rq3}

\input{sec/results/rq4}

%% file: sec/results/rq1.tex
\subsection{RQ1: {Detection Capability}}
\noindent \textbf{Experiment setup}. To answer this question, we directly run these five bug detectors on the four ML libraries, we record all the reported bugs (including the detailed location and CWE information) by these tools for further analysis. For each of the reported bugs, we further manually check whether it's a true bug or a false positive.

\input{tables/cweDistributionLib}



Table \ref{tab:cwelibdistribution} displays {the number of bug types extracted from warnings reported by each bug detector} across the four ML libraries. Specifically, among the four ML libraries, TensorFlow has the highest number of reported warnings, i.e., in total 445 warnings
\textit{Mlpack} has the lowest number of reported warnings overall, with only 26 warnings across all tools. Among the detectors, \textit{Clang static analyzer} reports the highest number of warnings for each library. \textit{Infer} only reports six warnings.

We further illustrate the top 10 bug types and their distribution in our curated dataset in Figure \ref{fig:typeDist}. The chart shows that the most common bug type is CWE-190 (Integer Overflow or Wraparound), with 86 occurrences in the studied ML libraries. The next most common types are CWE-362 (Race Condition), CWE-401 (Missing Release of Resource after Effective Lifetime), and CWE-476 (NULL Pointer Dereference), with 45, 43, and 34 occurrences, respectively. The least common bug type in the chart is CWE-122 (Heap-based Buffer Overflow), with only 16 occurrences.

Table \ref{tab:CWEDistribution} {shows the number of occurrences of specific CWEs reported by each tool and the false positives of these tools. The table also shows the ground-truth information about each CWE type in our curated dataset. Note that, some ground-truth CWEs are shown with (-), which means we do not have that specific type of bug in our dataset while the tools have generated warnings with that specific type.} 
For example, Flawfinder reported 50 instances of CWE-120 on these four ML libraries while 46 of them are false positives. In addition, all these reported 50 instances are outside of our experiment dataset. 
Overall, the five static analysis tools reported 138 instances of various types of CWEs in the studied ML libraries. Flawfinder reports the highest number of CWE types (85) while Infer reports the lowest (5). The most common CWE types are CWE-120 (classic buffer overflow) and CWE-121 (Stack-based Buffer Overflow), both of which are reported by Flawfinder and {RATS}. Other types of CWEs, such as CWE-362 (Concurrent Execution using Shared Resource with Improper Synchronization ('Race Condition')) and CWE-122 (Heap-based Buffer Overflow), are reported by only a subset of the tools. 

\input{figures/fig_vul_dist}

\begin{tcolorbox}
\textbf{Finding 1:} 
The number of warnings produced by the five static bug detectors varies significantly and is mainly determined by their detection mechanisms. While these tools can also generate false positives, and that human review is still necessary to determine whether reported warnings are real-world bugs that require attention.
\end{tcolorbox}


\input{tables/cweDistribution}

%% file: tables/cweDistributionLib.tex
\begin{table}[t!]
\centering
\caption{Distribution of reported warnings across different ML libraries.}
\label{tab:cwelibdistribution}
\vspace{-0.1in}
\setlength{\tabcolsep}{5pt}
\renewcommand{\arraystretch}{0.7}
\resizebox{\columnwidth}{!}{
\begin{tabular}{lccccc}
\toprule
Library     & Flawfinder & RATS & Cppcheck & Infer  & Clang static analyzer\\
\midrule
Mlpack      & 3          & 22   & 1        &  0     & 0   \\
MXNet       & 10         & 6    & 7        & 0       & 0  \\
PyTorch     & 5          & 1    & 2    & 4     & 16\\
TensorFlow  & 50        & 139  & 0       & 2       & 254   \\ \hline
Overall     & 68       & 168  & 10      & 6     & 270 \\
\bottomrule
\end{tabular}}
\end{table}

%% file: figures/fig_vul_dist.tex
\begin{figure}[t!]
\centering
\begin{tikzpicture}[scale=0.6]
\begin{axis}  
[  
    ybar,  
    enlargelimits=0.15,
    height=4.5cm,
    width=14cm,
    ylabel={\# of vulnerabilities}, 
    xlabel={CWE identifier},  
    symbolic x coords={190, 362, 401, 476, 908, 835, 121, 1339, 125, 122}, 
    xtick=data,  
     nodes near coords,
    nodes near coords align={vertical},  
    ]  
\addplot coordinates {
(190, 86)
(362, 45)
(401, 43)
(476, 34)
(908, 27)
(835, 26)
(121, 25)
(1339, 18)
(125, 17)
(122, 16)
};  
\end{axis}  
\end{tikzpicture}
\caption{The top 10 vulnerability types in our dataset.} 
\label{fig:typeDist}
\end{figure}
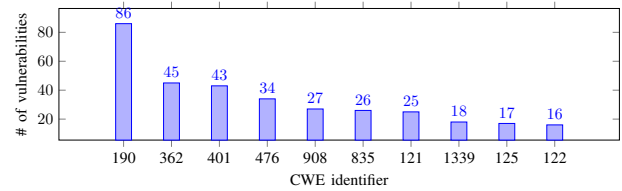

%% file: tables/cweDistribution.tex

\begin{table}[t!]
\centering
\caption{Distribution of vulnerability types reported by the four static vulnerability detectors. The warnings generated by Clang static analyzer do not have a CWE number, so we exclude it in this table. Numbers in the brackets are false positive warnings. The numbers outside of the brackets are the total number of generated warnings. } 
\label{tab:CWEDistribution}
\setlength{\tabcolsep}{5pt}
\renewcommand{\arraystretch}{1}
\resizebox{\columnwidth}{!}{
\begin{tabular}{lcccccc}
\toprule
CWE-ID  & \# True Bugs    & Flawfinder & RATS & Cppcheck & Infer & Overall \\
\midrule
CWE-120  & -  & 50 (46)         &   0    &  0        &   0    & 50          \\
CWE-121  &25    &    0         & 27 (26)   &   0       & 0    & 27          \\
CWE-122  &24   &    0            & 8 (5)    &    0      & 1 (1)    & 9           \\
CWE-362  &45   & 7 (7)              & 2 (2)   &    0     &    0   & 9           \\
CWE-327  &-   & 7 (7)             &   0   &    0      &   0    & 7           \\
CWE-367  &-  & 7 (7)           &   0   &   0      &   0    & 7           \\
CWE-20   &7  & 5 (4)           &   0   &   0      &   0     & 5           \\
CWE-476  & 34   &   0          &   0   & 5 (5)       &    0    & 5           \\
CWE-807  &-  & 4 (4)         &   0    &    0       &   0      & 4           \\
CWE-126  &-  & 3 (3)          &   0  &   0       &  0     & 3           \\
CWE-330 & -  &   0          & 3 (3)   &   0    &  0    & 3           \\
CWE-457  &1   &   0           &  0    &  0       & 2 (2)    & 2           \\
CWE-563  &-  &   0         &   0  &    0     & 1 (1)    & 1           \\
CWE-667  &-  &  0        &  0   & 1 (1)        &   0     & 1           \\
CWE-758  &- &    0        &  0    & 1 (1)       &  0     & 1           \\
CWE-190  &86  & 1 (0)          &  0    & 0        &  0     & 1           \\
CWE-398  &-  &  0         &  0   & 1 (1)       &   0    & 1           \\
CWE-119  &-  & 1 (1)        &  0   &    0     &  0    & 1           \\
CWE-833  &15  &    0        &  0   &    0     & 1 (1)    & 1           \\
CWE-401  &43  &       0      &  0    &     0     & 0   & 0  \\
CWE-908  &27  &       0      &  0    &     0     & 0   & 0  \\
CWE-835  &26  &       0      &  0    &     0     & 0   & 0  \\
CWE-1331 & 18   &       0      &  0    &     0     & 0   & 0  \\
CWE-125  &17   &       0      &  0    &     0     & 0   & 0  \\
CWE-369  &10  &       0      &  0    &     0     & 0   & 0  \\
CWE-703  &4   &       0      &  0    &     0     & 0   & 0  \\
CWE-191  &3   &       0      &  0    &     0     & 0   & 0  \\
CWE-705  &3  &       0      &  0    &     0     & 0   & 0  \\
CWE-415  & 3  &       0      &  0    &     0     & 0   & 0  \\
CWE-704   &2  &       0      &  0    &     0     & 0   & 0  \\
CWE-416   &2  &       0      &  0    &     0     & 0   & 0  \\
CWE-840   &2  &       0      &  0    &     0     & 0   & 0  \\
CWE-787   & 2  &       0      &  0    &     0     & 0   & 0  \\
CWE-439   & 2 &       0      &  0    &     0     & 0   & 0  \\
CWE-628   & 1  &       0      &  0    &     0     & 0   & 0  \\
CWE-241   & 1  &       0      &  0    &     0     & 0   & 0  \\
CWE-255    &1  &       0      &  0    &     0     & 0   & 0  \\
CWE-197    & 1  &        0      &  0    &     0     & 0   & 0  \\
CWE-252    &  1  &       0      &  0    &     0     & 0   & 0  \\
CWE-706     &  1  &       0      &  0    &     0     & 0   & 0  \\
CWE-1006   &   1  &       0      &  0    &     0     & 0   & 0  \\
CWE-561     &  1  &       0      &  0    &     0     & 0   & 0 \\
CWE-475     &  1  &       0      &  0    &     0     & 0   & 0  \\
\hline
Overall & 410 & 85                 & 40         & 8              & 5     & 138 \\
\bottomrule
\end{tabular}}
\end{table}

%% file: sec/results/rq3.tex
\subsection{RQ2: Detection Effectiveness}

\noindent \textbf{Experiment setup}.
To address this question, first two authors of this paper manually reviewed the generated warnings (bug candidates identified using the automatic filtering approach explained in subsection \ref{sec:mapping}) per file for each bug fixing commit. The team conducted a thorough analysis of the warning message, the CWE-ID, and the specific line numbers that can indicate where the bug was identified. Using this information, they further examined the code modifications to verify that the warning is indeed related to the bug being addressed in the particular update. If any discrepancies arise, they highlighted the corresponding warning and committed to additional examination in the subsequent review cycle. This process is repeated until all the warnings and code changes have been assessed. 

\noindent \textbf{Results}. 
Table~\ref{tab:mainPerf} shows the performance of each static bug detector regarding True Positive Rate (TPR) and False Negative Rate (FNR) values.  
As we can see from the table, Flawfinder has a TPR of 0.049, which means it correctly identifies only 4.95\% of the actual bugs, and an FNR of 0.95, indicating that it fails to detect 95.05\% of actual bugs. RATS has a TPR of 0.030 meaning that it correctly identifies only 3.07\% of the actual bugs, and a FNR of 0.97, which means it fails to detect 97.01\% of the actual bugs. The rest of the tools have a TPR of 0 which states that they fail to detect any actual bugs, and a FNR of 1, which means they incorrectly identifies all actual bugs as negatives.
Overall, all the experimented tools have relatively low performance in terms of identifying actual bugs in ML libraries, as they all have very low TPR values or fail to detect any real-world bugs at all.

\input{tables/tprfnrTable}
\input{tables/detailedTrueBugsDetected}

Table \ref{tab:detailedTrueBugs} shows the detailed characteristics of bugs detected by Flawfinder and RATS. 
We can see that Flawfinder and RATS can identify  bugs in different code repositories, including MXNet, TensorFlow, and PyTorch. The bugs are categorized according to their types ( i.e., CWE numbers). For example, Flawfinder detected a CWE-122 bug and a CWE-190 bug in files of MXNet. 
RATS also detected a CWE-122 bug in the same file of MXNet, as well as a CWE-190 bug in PyTorch. 



Table~\ref{tab:CWEDistribution} also shows the number of false positive warnings out of all warnings generated given each true bug type. 
Specifically, Flawfinder reported a total of 50 warnings for CWE-120, but only 4 of them were accurate warnings. In addition, it reported a single warning for CWE-190 that was indeed accurate. For CWE-20, Flawfinder reported 5 warnings, but only 1 of them was a true positive. On the other hand, RATS produced 27 warnings for CWE-121, but only 1 of them was a valid warning. For CWE-122, RATS flagged 5 warnings, but all of them were false positives. All warnings generated by Cppcheck and Infer are false positives suggesting that they are not effective at detecting real-world security bugs in the studied ML libraries. 




\begin{tcolorbox}
\textbf{Finding 2:} Overall, the effectiveness of the tools is quite poor in discovering real-world ML software bugs. Flawfinder and RATS, which are the most effective static checker, discovered 4 unique bugs out of a total of 410 bugs in our dataset.
\end{tcolorbox}



%% file: tables/tprfnrTable.tex
\begin{table}[t!]
\centering
\caption{Performance of the studied static vulnerability detectors on the four ML libraries.}
\label{tab:mainPerf}
\begin{tabular}{lcc}
\toprule
Tool       & TPR           & FNR          \\
\midrule
Flawfinder & 0.04 & 0.95 \\
RATS       & 0.03 & 0.97 \\
Cppcheck   & 0             & 1            \\
Infer      & 0             & 1            \\
Clang static analyzer & 0             & 1            \\
\bottomrule
\end{tabular}
\end{table}

%% file: tables/detailedTrueBugsDetected.tex
\begin{table*}[t!]
\centering
\caption{Vulnerabilities detected by Flawfinder and RATS.}
\label{tab:detailedTrueBugs}
\begin{tabular}{lccll}
\toprule
Tool  & Manifestation (ground-truth) & Root cause (reported by tools) & Library & Filename \\
\midrule
Flawfinder  & CWE-125 & CWE-20 &    PyTorch      &  \textit{simd.h}  \\
Flawfinder  & CWE-197 & CWE-122 &    PyTorch      &  \textit{decode\_padded\_raw\_op.cc}   \\
Flawfinder  & CWE-122 &  CWE-20 &   MXNet      &  \textit{image-classification-predict.cc} \\
Flawfinder    & CWE-190  & CWE-190 &  MXNet  &  \textit{image\_iter\_common.h} \\
Flawfinder    &  CWE-121 &  CWE-121 &  TensorFlow    &  \textit{cpu\_device.cc} \\
RATS  & CWE-122 &  CWE-121 &   MXNet   &  \textit{image-classification-predict.cc}  \\
RATS  & CWE-190  & CWE-122 &      PyTorch       & \textit{THTensor.cpp} \\
RATS  & CWE-197 & CWE-122 &     PyTorch      &  \textit{decode\_padded\_raw\_op.cc}  \\
\bottomrule
\end{tabular}
\end{table*}






%% file: sec/results/rq4.tex
\subsection{RQ3: Root Cause of Missing Real-world Bugs}
To address this research question, we manually analyzed and reviewed the documentation of tools as well as their code base. Then, we organized the reasons why these tools miss detect so many ML bugs.

\subsubsection{Flawfinder and RATS} 
Since Flawfinder and RATS manifest very similarly in bug detection, Flawfinder's specific reasons also apply to RATS. 

\noindent \textbf{Issues with Soundness Strategy.} {Flawfinder and RATS have a significant limitation in their approach to bug detection, known as the soundness strategy. This strategy leads them to flag every C/C++ API call as potentially buggy, which is a major drawback. Although ML libraries often make extensive use of C/C++ APIs in their backend implementation, the bugs in these libraries do not primarily originate from these APIs. Instead, the bugs in ML libraries tend to exhibit more complex patterns (require a deeper source code analysis in order to be detected), which are not effectively captured by the simplistic approach of flagging all API calls as} buggy~\cite{harzevili2022characterizing, islam2019comprehensive, thung2012empirical, xiao2018security, thung2012empirical, jia2021symptoms, di2017comprehensive, shen2021comprehensive}.  

\begin{figure}[t!]
\centering
    \begin{lstlisting} [
                        escapechar=`,
                        numbers=left,
                        firstnumber=84,
                       language = C,
                       escapeinside={(*}{*)}]
auto input_shape = c->input(0);
auto input_h_shape = c->input(1);
auto seq_length = c->Dim(input_shape, 0);
// assumes rank >= 2
auto batch_size = c->Dim(input_shape, 1); 
// assumes rank >= 3
auto num_units = c->Dim(input_h_shape, 2); 
    \end{lstlisting} 
  \caption{An example of heap buffer overflow in TensorFlow library.}
\label{fig:tensorheaperror}
\end{figure}

\noindent \textbf{Lack of ML-specific Buggy API Information.} 
{Even though Flawfinder and RATS support more than 200 dangerous C/C++ APIs, they are still incapable of supporting ML-specific buggy APIs. One of the reasons they miss detecting real-world security bugs in ML libraries is that they do not model library-specific API} information~\cite{harzevili2022characterizing, jia2021symptoms}. For example, to detect \textit{Memory Leak (CWE-401)} in the Mlpack library, the static detectors need to model \verb|CleanMemory()| or \verb|delete_mat| which are Mlpack-specific APIs used for cleaning allocated memories.

\vspace{4pt}
\noindent \textbf{Lack of Data Flow and Control Flow Support.} 
{Flawfinder and RATS face challenges in detecting numerous bug patterns specific to machine learning (ML) because they lack the capability to model control flow and data flow. Instead, these tools rely on a simplistic keyword-matching approach to identify potentially dangerous C/C++ API calls and flag them as bugs. This limitation hinders their effectiveness in capturing the intricacies of ML-specific bugs~\cite{jia2021symptoms, harzevili2022characterizing} that involve complex control and data dependencies. For example, lack of validation is a complex root cause pattern that is the major root cause of data type bugs in ML libraries~\cite{harzevili2022characterizing}. In this particular bug pattern, bugs can arise when utilizing client APIs of ML libraries. If the backend implementation fails to validate or properly handle malicious inputs or malformed values received through these APIs, bugs can occur. It is crucial to perform appropriate validation and handling of such inputs to mitigate potential issues and ensure the robustness and security of the ML library. For example}, Figure~\ref{fig:tensorheaperror} is an example of a lack of validation that causes a heap buffer overflow. In this bug, the code assumes \verb|input_shape| and \verb|input_h_shape| have a specific rank, while the rank values should be validated to avoid possible invalid memory access. Flawfinder and RATS are not able to detect such bugs because they are incapable of modeling data flow dependency and keeping track of \verb|input_shape| and \verb|input_h_shape| to check if these variables are validated or not.

\subsubsection{Cppcheck}
\noindent \textbf{Limited Buffer or Stack Overflow Checking}
{Existing work~\cite{harzevili2022characterizing} has shown that the root cause patterns of buffer overflow or stack overflow in ML libraries are significantly different compared to that of traditional software. 
For example, in this commit\footnote{\tiny{https://github.com/tensorflow/tensorflow/commit/932c4c2364884af52609ea8a86c7232a926d958f}} from the TensorFlow library, the root cause of stack overflow is a very big computation graph of functions and edges.} In this example, Each instance of the class \verb|std::shared_ptr<Function>| holds a collection of Edge objects, and each Edge object, in turn, holds a \verb|std::shared_ptr<Function>|. Removing a \verb|std::shared_ptr<Function>| can lead to the cascading deletion of other \verb|std::shared_ptr<Function>| instances, potentially resulting in a stack overflow if the graph has a significant depth. However, Cppcheck performs different strategies to detect buffer overflow and stack overflow bugs which is not strong enough to detect ML-related overflow bugs. For example, Cppcheck uses array index checkers to detect buffer overflow which identifies array index operations and performs various checks to detect potential buffer overruns associated with array indexing. More specifically, Cppcheck exhaustively searches for array indexing statements in the code snippet, even if the code is not reachable, without any control flow analysis. This technique is incapable of detecting overflow bugs which is due to the large computation graph mentioned above. The second technique to detect buffer overflow or stack overflow is to search for C/C++ API calls, similar to Flawfinder and RATS explained earlier. In this technique, Cppcheck examines the API scopes and the corresponding arguments, analyzes the buffer or stack sizes associated with the arguments, and detects potential buffer overflow or stack overflow bugs based on specified minimum size requirements.

\noindent \textbf{Limited Control Flow Analysis}
One major limitation of Cppcheck is limited control flow analysis, while a strong control flow analysis is required to detect very hard-to-detect bugs In ML libraries, e.g., memory leaks that have complicated bug patterns. For example, this memory leak bug\footnote{\tiny{https://cve.mitre.org/cgi-bin/cvename.cgi?name=CVE-2022-23585}} occurs in the TensorFlow library when decoding malformed PNG image. The memory leak occurs when certain errors in the function implementation cause the execution to be abruptly terminated using the \verb|OP_REQUIRES| macro checker. This termination prevents the proper freeing of allocated buffers stored in the \verb|decode| value. To release these allocated buffers, the function should call \verb|png::CommonFreeDecode(&decode)|. However, due to eager termination, the necessary memory-freeing process is not allowed to occur. This bug is very hard to detect by Cppcheck for two main reasons. First, the TensorFlow-specific API \verb|png::CommonFreeDecode(&decode);| is not registered in the internal database of API symbols which is the firsts step toward detecting this leak. Second, the control-flow analysis supported by Cppcheck is very simple and cannot capture the complex flow analysis in this bug pattern.

\subsubsection{Infer}

\noindent \textbf{Limited Buffer Overflow checkers}
Compared to Cppcheck, Flawfinder, and RATS, Infer has a stronger buffer overflow checker. Infer uses symbolic intervals to handle the range of index values and buffer sizes. Typically, interval analysis involves working with intervals represented as [low, high], where low and high are constants indicating the lower bound and upper bound of the target buffer and indexing range. 
However, this checker is not effective enough to detect ML-related buffer overflow bugs. For example, Figure~\ref{fig:tensorflowbufferoverrun} shows an example of hard-to-detect buffer overflow\footnote{\tiny{https://github.com/tensorflow/tensorflow/commit/8ee24e7949a203d234489f9da2c5bf45a7d5157d}} in the TensorFlow library where providing a tensor with larger dimensions as the second argument is critical. To address this issue, the developer has modified the \verb|MatchingDim| function to return the minimum size between the two dimensions. 

\begin{figure}[t!]
\centering
    \begin{lstlisting} [
                        escapechar=`,
                        numbers=left,
                        firstnumber=84,
                       language = C,
                       escapeinside={(*}{*)}]
inline int MatchingDim(const RuntimeShape& 
shape1, int index1, const RuntimeShape& 
shape2, int index2) {
  TFLITE_DCHECK_EQ(shape1.Dims(index1), 
  shape2.Dims(index2));
  return shape1.Dims(index1);
}
    \end{lstlisting} 
  \caption{An example of hard-to-detect buffer overflow bug in TensorFlow library.}
\label{fig:tensorflowbufferoverrun}
\end{figure}



\subsubsection{Clang static analyzer}

\noindent \textbf{Limited Rules}. { Although Clang static analyzer provides a large number of predefined rules (i.e., there are 25 families of rules implemented in the Clang static analyzer database), it does not cover rules of security bugs that are relevant to ML libraries. Most of the rules are implemented to find coding conventions and styling errors such as \textit{fuchsia\_} which check the correctness of codes based on \textit{Fuchsia} coding conventions. There are also two broad categories of rules for performance and concurrency issues. However, none of them are specifically designed to find security issues for ML libraries. Concurrency issues in ML libraries have very sophisticated bug patterns \cite{harzevili2022characterizing} that require customized ML-specific rules to be implemented in Clang static analyzer.} 

\vspace{4PT}
\noindent \textbf{Limited Sensitivity}. 
{Even though the Clang static analyzer supports all sensitivity analysis factors explained in Table~\ref{tab:toolcomparison}, there are still some major issues regarding sensitivity analysis for ML libraries. For example, it may generate false positive warnings suggesting that a variable is uninitialized while it is properly initialized. This is because flow analysis in Clang static analyzer often is not performed efficiently. This flow analysis support may not be able to catch the bug pattern of uninitialized variables in ML libraries. Additionally, it may not be able to cover all cases of uninitialized variables as there may have sophisticated patterns in ML libraries implementation. }

\begin{tcolorbox}
\textbf{Finding 3:}
We identified a set of specific reasons that help explain why the five static bug detectors examined in this paper miss real-world security bugs in ML libraries including ML software-specific reasons, i.e., \textbf{Lack of Implementation Support of Detection Rules}, \textbf{Issues with Soundness Strategy}, and \textbf{Lack of ML-specific Buggy API Information}.


\end{tcolorbox}

%% file: sec/lessons.tex
\section{Lessons learned}

Our study reveals several interesting findings that can serve as applicable guidelines for improving static bug detection for ML libraries.

\subsection{Implication for Improving Flawfinder and RATS}
We find that Flawfinder and RATS flag every C/C++ API as vulnerable regardless of whether they are actually vulnerable or not. This behavior introduces many false alarms \ref{tab:CWEDistribution}. In order to reduce the false alarm rate, numerous extensions to these checkers are required. 


\vspace{4pt}
\noindent \textbf{Extend the input representation.}
Flawfinder and RATS represent the source code as a sequence of tokens. Unlike natural language, source code encodes structural information \cite{Hindle}, which needs to be considered with rich representation techniques like Abstract Syntax Trees(AST), Control Flow Graphs(CFG), or Data Flow Graphs(DFG)~\cite{yamaguchi2014modeling}. Having a control flow graph is vital in detecting many ML-related security vulnerabilities including buffer overflow and memory leaks. Control flow graphs allow the tool to model every possible execution patch inside vulnerable programs. For example, to detect the stack overflow bug in this commit \footnote{\tiny{https://github.com/tensorflow/tensorflow/commit/698bc996f7190f5cd836d48d29b8c1b3ddcd37c2}}, the checker needs to traverse control graph to find out that the implementation of \verb|DataTypeString| function is vulnerable due to missing \verb|return DataTypeStringInternal(dtype);|. 

\vspace{4pt}
\noindent \textbf{Extend the pattern matching.}
The pattern-matching strategy of Flawfinder and RATS follows naive text-based matching. Text-based matching does not allow the detectors to perform intraprocedural analysis, i.e. the analysis inside the boundary of functions or any compilation units. The lack of performing intraprocedural analysis introduces so many false alarms in the case of detecting ML security vulnerabilities. For example, in order to detect \textit{Memory Leaks (CWE-401)} bug listed in Table \ref{tab:detailedTrueBugs}, the detector needs to perform analysis inside \verb|_bsplmat()| and catch the return value which is the root cause of this bug. 

\vspace{4pt}
\subsection{Implication for Improving Cppcheck}
Cppcheck has been shown to be the most effective static checker for detecting real-world security issues. However, it still lacks a plethora of weaknesses. Cppcheck may be extended in two ways, as discussed in the subsections below. 

\vspace{4pt}
\noindent \textbf{Support ML Library-Specific Constructs.}
In order to extend Cppcheck, the developers should handle any ML library-specific constructs or macros within the backend implementation. This may involve extending Cppcheck's macro handling capabilities or creating additional checks to handle these constructs effectively.

\noindent \textbf{Extend the control flow graph.}
The control flow graph analysis in Cppcheck is very limited, with the following assumptions: all source code statements can be accessed, and the state checkers in the if conditions are always either true or false. As a result, a more advanced control flow graph is required to be able to detect sophisticated vulnerabilities in ML libraries.

\subsection{Implication for Improving Infer}

\noindent \textbf{Improve Buffer overrun Checker.}
While Infer's buffer overrun checker can detect typical buffer overflow patterns, it may not cover all conceivable variations and attacker approaches in ML libraries. The tool may miss sophisticated or unique exploitation methods explained in RQ3.

\noindent \textbf{Isolate checkers.}
To begin with, the rules for identifying integer overflow in Infer are still in an experimental stage and lack sensitivity to function arguments. Additionally, Infer's dependency on buffer overrun rules to detect integer overflow introduces a limitation. These buffer overrun rules themselves tend to generate a significant number of false alarms, which consequently raises the false alarm rate for identifying integer overflow bugs. Given the intricate nature of the bug pattern for integer overflow in ML libraries~\cite{harzevili2022characterizing}, it becomes crucial for developers to separate the buffer overrun checkers from the scope of integer overflow.

\subsection{Implication for Improving Clang static analyzer}

\noindent \textbf{Extend the Checker for Null Pointer Dereference}.
Clang static analyzer uses its internal checker \textit{core.NullDereference (C, C++, ObjC)} to detect null pointer dereference bugs in C, C++, and Objective C programs. The checker works well if function arguments take pointer parameters. If the code attempts to dereference the pointer without checking if it is null, it marks it as buggy. However, in terms of ML libraries, null pointer dereference has more sophisticated patterns. For example, in this commit from the TensorFlow library\footnote{\tiny{https://github.com/tensorflow/tensorflow/commit/9a133d73ae4b4664d22bd1aa6d654fec13c52ee1}}, the developer has removed \verb|int64 id = ctx->session_state()->GetNewId();| since \verb|session_state()| is vulnerable to have a null value which results in denial of service via a null pointer dereference. The developers should increase the null pointer checker to cover more corner cases in terms of code elements that may have null values.

%% file: sec/validity.tex
\section{Threats to validity}


As is the case with every empirical research, there are a few factors that call into question the reliability of the inferences we have taken from our data. One of the major limitations is the choice of ML libraries and bug detectors. In order to protect ourselves against this risk, we have chosen four extensively used ML libraries, each of which focuses on a different facet of ML development. All ML libraries are open-source projects that are always being developed and improved upon. Regarding the selection of static bug detectors, our major focus is on static detectors that are openly available to the public and are in the process of being developed right now. We used widely used and popular detectors which have been cited by many previous studies in the bug detection areas ~\cite{cheng2021deepwukong, li2018vuldeepecker, duan2019vulsniper}.

Another possible threat to this study is the mapping technique \cite{habib2018many} used to find potential vulnerable candidates. The approach we used in the paper is subject to coincidental matches. for example, the assumption is that if the line number produced in warnings overlaps with the modified lines in a bug fix, the program will automatically recognize the warning as a candidate for the bug. This is not true in practice since the reported warning may not be connected to the actual bug in the commit, or the changed line may be refactoring the code and does not represent the bug being fixed. The ultimate determination of whether a warning relates to a bug is made by two authors involved in manual inspection and is therefore subjective. To mitigate this risk, both authors reviewed every possibility for an identified bug when there is no clear evidence. 

In this paper, we rely on the assumption that a source code before a bug fix is a buggy source code, and the source code after the fix is considered as the bug is fixed. One may apply static bug detectors on source code after the fix and find multiple bugs. As a result, this assumption serves as the foundation for this paper's goal of determining how many real-world bugs the tools identify at the moment of committing modifications. 

%% file: sec/related.tex
\section{Related work}

\subsection{Software Security Vulnerability Detection}
There have been numerous studies focused on software vulnerability detection in the literature \cite{cao2022mvd, li2021sysevr, cheng2021deepwukong, duan2019vulsniper, li2018vuldeepecker, li2021vuldeelocator}. Cao et al.~\cite{cao2022mvd} proposed a deep learning-based vulnerability detection model to detect memory-related statements on their manually curated dataset extracted from 11 projects developed in C/C++. Their proposed model can model structural information which allows the detection of semantic vulnerabilities. Their experiments indicate that the proposed model is superior compared to cutting-edge models as well as static analysis tools. Li et al.~\cite{li2021sysevr} developed SySeVR, a deep learning-based vulnerability detector in which syntactic and semantic information contained in source codes are merged as a rich input representation and supplied into the model. They think that this unique representation discovers subtle weaknesses in the source code. SySeVR surpasses state-of-the-art models and static checkers in detecting many real-world security vulnerabilities. Cheng et al \cite{cheng2021deepwukong} focused on combining static detection and deep learning models to detect software security vulnerabilities. They proposed a novel embedding framework called DeepWukong in which a graph neural network is used to model semantic code logic via control and data flow analysis. They gathered more than 100,000 programs written in C/C++ to evaluate DeepWukon's capabilities for detecting real-world software security vulnerabilities. They find that DeepWukong outperforms existing state-of-the-art models and static analysis tools in finding real-world security vulnerabilities. 

\subsection{Studies on Static Detection Techniques}

Habib and Pradel \cite{habib2018many} investigated the static analyzers including Infer, ErrorProne, and SpotBugs to figure out what percentage of Defect4j's total bugs can be located by using the aforementioned tools. Both the code diff and the bug report mapping approaches are utilized by the authors. Tomassi \cite{tomassi2018bugs} They carry out a study in which they examine the similarities and differences between ErrorProne and SpotBugs in order to determine the total number of bugs that are discovered in a sample of 320 BugSWARM artifacts. SpotBugs was only able to locate a single bug, as the author discovered. 
Rutar et al. \cite{rutar2004comparison} analyzed a small suite of programs with a number of different static analyzers, including PMD, FindBugs, JLint, Bandera, and ESC/Java 2. The authors present a taxonomy of bugs discovered by each tool, demonstrating that none of the tools can be considered to be more comprehensive than the others.
Runtime as well as the total number of warnings generated are the primary focuses of this study. 
Chatzieleftheriou and Katsaros \cite{chatzieleftheriou2011test} focused on common static analysis techniques for C code security bugs. they proposed a test suite comprised of common bugs drawn from public databases using four open-source and two commercial projects. They proposed a set of findings that assist in the identification of cost-effective tools, and the proposed methodology and test suite can be used to evaluate static analysis tools in the future. 
Lipp et al. \cite{lipp2022empirical} argued that there are two main limitations to the datasets used for testing static bug detectors. Firstly, the datasets do not accurately represent real-world code and cannot identify new and complex vulnerability patterns. Secondly, the datasets do not classify vulnerabilities based on the Common Weakness Enumeration (CWE) mapping. Hence, they attempted to overcome these limitations by testing static bug detectors on real-world datasets that were collected from CVE records of 27 different projects, totaling 1.15 million lines of code. In spite of the fact that static bug detectors perform well on the synthetic dataset of bugs, they find that the tools miss a high percentage of bugs in a real-world setting.



Static bug detectors often produce a large number of false alarm warnings which makes manual inspection problematic. To remove such false alarms,  Several different approaches to the detection of static bugs have been taken up by the industry. Bessey et al. \cite{bessey2010few} share their insights gained from the process of bringing static bug detectors to the market in their paper. Ayewah et al.~\cite{ayewah2008using, ayewah2010google} discussed the lessons learned via applying FindBug on Google's codebase in which numerous engineers involved through thousands of FindBugs generated warnings, and addressed them by either fixing them or filing reports. They find that most issues were highlighted for fixing, but only a few of them were actually causing significant problems in production.

Understanding real-world bugs is a crucial first step in enhancing bug detection. Several studies have taken different types of bugs into account, such as those in the Linux kernel \cite{chou2001empirical}, bugs in concurrency \cite{lu2008learning}, and bugs in correctness \cite{ocariza2013empirical} and performance \cite{selakovic2016performance} in JavaScript. The work of Pan et al. cited as \cite{pan2009toward}, examines fixes for bugs and identifies recurring syntactic patterns. 

Our work is very different from the studies that have been done before \cite{lipp2022empirical,habib2018many, tomassi2018bugs, tomassi2021real}.
In previous studies, the primary focus has been on the automatic static detection of general bugs in traditional software projects. On the other hand, the primary focus of this paper is on the automatic detection of software bugs in widely used ML projects. For example, Integer Overflow (CWE-190) is a major bug in ML libraries as found in~\cite{harzevili2022characterizing}. A further distinction lies in the fact that the focus of our research is placed on a more recent and advanced generation of static vulnerability detectors specifically designed to detect bugs in projects written in C/C++.

%% file: sec/conclusion.tex
\section{Conclusion}

This paper addresses the critical task of automatic detection of software bugs in ML libraries. The study analyzes the effectiveness of five popular and widely used static bug detectors, namely Flawfinder, RATS, Cppcheck, Facebook Infer, and Clang static analyzer, on a curated dataset of software bugs gathered from four popular ML libraries. The research provides a categorization of these tools' capabilities, highlighting their strengths and weaknesses in detecting software bugs. The study reveals that Flawfinder and RATS are the most effective static checkers for finding security bugs in ML libraries. However, the overall findings show that the tools detect only a negligible amount of bugs, accounting for 6/410 (0.01\%) of known security bugs. Based on these observations, the paper also identifies and discusses opportunities to make the tools more effective and practical.